**Does cross-modal discounting generalize to non-WEIRD cultures? A comparison of the USA and Japan**

Shohei Yamamoto[1], Rebecca McDonald[2] and Daniel Read[3]

Abstract

This paper examines how outcome modality in intertemporal choice influences time preferences and whether the process differs across cultures, specifically Japan and the United States. Uni-modal choices are those when the outcomes being compared over time are very similar, and cross-modal choices are those when the outcomes are very different. The cross-modal effect, previously shown in the U.S., is that there is greater patience in cross-modal decisions. In Experiment 1, we employed a between-participants design, in which participants either made uni-modal or cross-modal decisions. In Experiment 2, we employed a within-participants design in which everyone made both types of decision. In both Experiments we replicated the cross-modal effect. Moreover, the magnitude of the effect did not vary with factors known to relate to time preference, such as cognitive ability and social status, and it did not differ across cultures, even though Japanese participants were much more patient than American ones. The effect was stronger in the between- than within-participants experiment. These results strengthen the conclusion that the cross-modal effect is universal and strengthens the argument that it is due to the fundamental process of attentional dilution.

[1] Department of Economics, Rikkyo University, Tokyo, Japan Email: shohei.yamamoto@rikkyo.ac.jp ORCID 0000-0002-8136-6842

[2] Department of Economics, University of Birmingham, Birmingham, UK. Email: r.l.mcdonald@bham.ac.uk ORCID 0000-0003-1593-7675

[3] Corresponding author. Warwick Business School, University of Warwick, Coventry, UK. Email: daniel.read@wbs.ac.uk. ORCID 0000-0003-1593-7675

**Does cross-modal discounting generalize to non-WEIRD cultures? A comparison of the USA and Japan**

When offered a choice between a positive outcome now and the same outcome later, most people prefer the former and will require a better outcome later to induce them to wait. This is time discounting. As pointed out by economists since Fisher (1930; c.f., Mandler, 1999), some degree of time discounting is quite reasonable and even wholly rational. But from an economic perspective this discounting should be stable, meaning that the discounting of a given outcome should not depend on how it is assessed. But studies have revealed many "anomalies", meaning systematic deviations from the patterns of choice that would be expected if people strive to optimize their lifetime stream of consumption. Some anomalies, such as the magnitude and sign effects, show that people apply different discounting to what should be theoretically equivalent outcomes (e.g., Loewenstein & Prelec, 1992; Roelofsma, 1996; Thaler, 1981). Other anomalies are even more fundamental, in that the same outcome is discounted differently depending on how it is presented or described or how responses are elicited (e.g., Lempert & Phelps, 2016; Read et al., 2018; Tversky et al., 1988). These anomalies present clues to the deeper processes of decision-making.

A recently described anomaly of the latter sort is the cross-modal effect (Cubitt et al., 2018; Read et al., 2023): the desire for immediate relative to delayed consumption is reduced when intertemporal decisions are transformed from uni-modal to cross-modal. Uni-modal decisions are those when the sooner and later outcomes in an intertemporal choice are of the same kind. Deciding between an ordinary television now and a high-end television in one year is uni-modal. Cross-modal decisions are those when the outcomes being traded off are of different kinds. Deciding between a weekend getaway now and a high-end television in one year is cross-modal. The cross-modal effect suggests the television will be discounted less (i.e., have a higher present value) in the second case than the first.

Apart from their theoretical interest, cross-modal decisions are important because they reflect many real-world settings. A notorious and colorful illustration came from Tim Gurney, the Australian millionaire, who argued that young people were making the wrong choice between avocado toast and expensive coffee now versus a home of their own later (Levin, 2017). Although Gurney's quantitative claim was rightly mocked, lifestyle decisions often do involve tradeoffs between qualitatively different outcomes at different times. The decision to smoke (or not), for example, involves a trade-off between the pleasure of smoking now and enjoying good health in later life.

The cross-modal effect has only recently been discovered since most studies of intertemporal choice have focused on uni-modal preferences, typically of the particular kind in which the trade-offs are some money now and (more) money later, but also trade-offs between, for example, food and other consumer goods, and states of health (Cheung et al., 2022; Dong et al., 2016; Reuben et al., 2010; Ubfal, 2016). Reviews of this work can be found in Frederick et al. (2002), Urminsky and Zauberman (2015) and Read et al. (2018). The focus on uni-modal choices is partly driven by methodological convenience, since uni-modal time preference (or discount) rates are easy to estimate. This is done by establishing the magnitude of the payoffs obtained in options differing in delay such that the decision maker is indifferent between them and then computing the implied rate per unit time at which payoffs must improve to maintain indifference. For instance, if you are indifferent between 100 grams of chocolate now and 200 grams in one year, this implies the quantity of chocolate must increase at a rate of 100% per year to maintain indifference.[1]

---

[1] This 100% increase does not only capture time preference, but also diminishing marginal utility of grams of chocolate. For goods with high satiation, the utility of their consumption may quickly decline as quantity increases, increasing the quantity demanded to maintain indifference. This means that discounting measured in terms of the quantity of chocolate may exceed the (theoretically relevant) discounting of the utility of consuming chocolate. This is another explanation for the reliance on monetary payments in the intertemporal choice literature to date: utility for money is likely to be essentially linear over the ranges traditionally used in experiments.

Assessing cross-modal discounting, and comparing it to uni-modal discounting, is a more elaborate process than measuring a single discount rate. It involves the integration of four experimental conditions, each involving one type of intertemporal comparison. As explained at length in earlier papers (Cubitt et al., 2018; Read et al., 2023), these are needed to theoretically "cancel out" the effect of different goods on discounting. To illustrate, take four comparisons: outcomes A now versus A later, B now versus B later, A now versus B later, and B now versus A later. For each comparison, measure the amount of money (or any common currency) an agent needs to make them just willing to switch from the earlier to the later option (we call this the "compensation"). The first two compensations are uni-modal, the second two are cross-modal. By comparing the average uni-modal compensation to the average cross-modal compensation we can determine if people discount less, more, or by the same amount cross-modally as uni-modally. The logic is that because the cross-modal and uni-modal decisions involve the same intertemporal changes between goods, but differently matched, the effect of the goods will cancel out, leaving only the effect of time discounting even in the cross-modal condition.

Under the usual theoretical assumptions, discounting is constant for all outcomes and preferences do not depend on whether mutually exclusive outcomes differ in kind.[2] Consequently, whether a choice is uni-modal or cross-modal should not influence patience, as demonstrated by Cubitt et al. (2018). Contrary to this theoretical argument, and as already mentioned, Cubitt et al., 2018; see also Read et al., 2023) showed that the effect of delay on intertemporal tradeoffs is significantly smaller in cross-modal decisions. They attributed this to *attentional dilution*, arguing that the weight placed on time[3] in decision-making is reduced as the outcomes being traded off become more different. In cross-modal decisions attention is "diluted" because it is allocated over the differences between goods and the differences in the time of their receipt, while in uni-modal decisions almost all the attention is allocated to the difference in time. Accordingly, more attention, and subsequently more decision weight, is put on time in uni-modal decisions.

---

[2] This does not mean observed discounting will be independent of what is being discounted. Rather, discounting for a given outcome that will be received later will depend on the outcome and its delay, and not on outcomes that will not be received, such as the comparator in an intertemporal choice.

[3] Beyond just time, Read et al. (2023) argued that the impact of any attribute, including risk, would be reduced as the number of differentiating attributes increases.

A natural question concerns the universality of the cross-modal effect. Earlier studies were conducted exclusively with American participants. We know that time preferences, and the rate of time discounting itself, differ widely across cultures (e.g., Burro et al., 2022; Falk et al., 2018; Ruggeri et al., 2022; Wang et al., 2016). Apart from the mere rate of discounting, it is also possible that qualitative effects such as cross-modal discounting differ from country to country. This is relevant to recent work calling into question whether findings from "WEIRD" societies (Western, Educated, Industrialized, Rich, and Democratic) can be assumed to apply to non-WEIRD ones (Henrich et al., 2010; Pitesa & Gelfand, 2023). Henrich et al. (2010) found that WEIRD and non-WEIRD populations differ in many important aspects of behavior, emphasizing the need to consider non-WEIRD groups when establishing and generalizing patterns of behavior. Moreover, they demonstrate that many biases observed in WEIRD countries are not universal. In this paper we test whether we will observe the cross-modal effect in a non-WEIRD country, specifically Japan.

The choice of countries is theoretically grounded. It is recognized that there are cultural differences in time orientation between East Asia and North America (Gao, 2016, Hou et al., 2024). Two widely held stereotypes are that Americans are impatient and Japanese people are patient (e.g., Graham & Sano, 1984; Hofstede insights, n.d.; Levine, 2008). For instance, Lewis (2019) describes the concept of *gaman*, a Japanese cultural value of endurance, and contrasts it with the western desire for immediate gratification. Experimental studies have found a significant level of support for these contrasting images. In particular, individuals from Japan and other Eastern cultures typically discount future rewards less than their American or Western counterparts (Du et al., 2002; Takahashi et al., 2009; Ishii et al., 2017; Kim et al., 2012; Raineri et al., 2024).

Proposed mechanisms for this difference include cultural and linguistic factors. One cultural factor is individualism. Eastern countries, exemplified by Japan's collectivistic culture, tend to emphasize long-term orientation, potentially contributing to less discounting of future rewards (Wang et al., 2016). In contrast, Western countries like the United States, characterized by individualistic culture, may exhibit a stronger preference for immediate rewards and a higher degree of time discounting.

Language may also influence time preferences. Japanese, a weak future time reference (FTR) language, grammatically assimilates the future and present more closely than English, a strong FTR language. For example, in weak-FTR languages, "I go shopping" applies both when the shopping trip happens in the present and when it will happen in the future, with context supplying the temporal information. In Japanese, a weak-FTR language, the sentence 買い物に行く (*kaimono ni iku*) literally means "(I) go shopping" and can refer to either a present or future shopping trip. To indicate the future more explicitly, a time adverb such as 明日 (*ashita*, "tomorrow") can be added: 明日、買い物に行く ("Tomorrow, I go shopping"). A strong-FTR language like English or French, by contrast, linguistically differentiates these cases. English distinguishes "I am shopping" from "I will go shopping," and French distinguishes "*je fais du shopping*" from "*je vais faire du shopping.*" There is now substantial evidence that this linguistic difference is associated with time discounting levels across cultures (Ayres et al., 2023; Chen, 2013), with those using strong-FTR languages valuing future events less than those using weak-FTR ones. Whether this effect is causal or due to some other factor is yet to be determined, since experimental studies have been less clear (e.g., Chen et al., 2019; Keller et al., 2024).

The existing literature has emphasized a discrepancy in time discounting between Japanese and American people, suggesting that what we learn from WEIRD people, at least with respect to the level of discounting, need not generalize fully to other cultures. The attentional dilution hypothesis suggests that the cross-modal effect will be relatively constant across different cultures. Attention is a lower-level cognitive process, and so if attentional dilution is the mechanism underlying the cross-modal effect, then we would expect it not to vary from culture to culture once we account for the expected initial lower levels of discounting among Japanese participants (although see Hedden et al., 2008).

To test this, we replicated earlier studies of the cross-modal effect in the USA and Japan. We deployed a between-participants design, replicating earlier studies, and also a within-participants design, in which each participant made both uni-modal and cross-modal decisions, allowing us to analyze the cross-modal effect at the individual level. The latter conditions also permitted us to test, by comparing across studies, a further prediction of the attentional dilution mechanism: as time is increased in salience even in cross-modal conditions, the importance of delay will also increase.

### Experiment overview

Experiment 1 employed a between-participants design similar to Read et al. (2023). Participants were randomly assigned to one of four conditions: two uni-modal and two cross-modal. Experiment 2 was a within-participants design, where everyone took part in all four conditions in a randomized order.

The methods and analysis plan for both studies were pre-registered at aspredicted.org: Experiment 1 is #207920, Experiment 2 is #120031. The experiments were conducted in the opposite order of that described below, but we describe them this way because Experiment 1 is a very close cross-cultural replication (with small variations) of the between-participants design adopted earlier (Read et al., 2023), and Experiment 2 is a within-participants follow up. The Appendix describes two within-participants pretest experiments conducted in Japan only, which yield results highly comparable to the Japanese sample in both experiments. The discussion considers what we can learn from all studies completed to date.

We obtained data through Prolific (https://www.prolific.co/) for American participants and Lancers (https://www.lancers.jp/) for Japanese participants. Prolific is a widely used crowdsourcing platform with participants in many countries, primarily English-speaking ones, but at the time of writing had only 304 Japanese participants. Lancers is a Japan-only market research company. It is one of the largest in Japan and functions similarly to Amazon Mechanical Turk. It is widely used in academic research for recruiting Japanese participants with characteristics comparable to those of Prolific participants. All studies involved a unique sample.

Both experiments received ethical approval from Hitotsubashi University (No. 2022D009).

## Experiment 1

### Methods

#### *Participants.*

We aimed for a sample size of 800 per country (200 per condition). This target was determined based on a power analysis conducted prior to data collection. Using effect size estimates from two pretests (expected effect size = 0.30, SD = 1.00), we calculated that 700 participants (175 per group) would be needed to detect a meaningful treatment effect with adequate power. As pre-registered, we oversampled to account for potential exclusions, trying for 860 participants in the U.S. and 820 in Japan. The smaller request in Japan was because, based on prior experience, Japanese participants were expected to be subject to fewer exclusions based on attention and consistency. Lancers was only able to provide 745 participants before exclusions.[4] We obtained 855 participants from the United States, losing five who did not complete the survey.

We excluded participants who did not meet one or more pre-registered criteria: (1) a consistency score below 0.5, (2) providing the wrong answer to a simple attention check, and (3) the selection of an extreme response in one or both of two test questions (described shortly). The consistency score, as described in Read et al. (2023), is the proportion of a participant's choices consistent with a single discount rate. A consistency score below 0.5 therefore strongly suggests random responding. The attention check was a standard one, provided by Qualtrics (who also tested it – Geisen, 2022), which involved identifying the vegetable from a list of foods. In the U.S. sample, four participants had a consistency score below 0.5, six got the attention check incorrect, and 79 selected extreme choices, with some failing on more than one criterion. In the Japanese sample, 44 selected extreme choices, and three had a consistency score below 0.5. No Japanese participant failed the attention check.

Following exclusions the sample consisted of 768 American participants (49.7% female; *M age* = 38.7 years, *range* = 18–80) and 701 Japanese participants (48.1% female; *M age* = 43.8 years, *range* = 18–75). While their average ages differ slightly, both samples were close to the mean age for the general populations in the two countries (38.9 USA, 49.9 Japan; WorldData.info, 2025). American participants received a fixed participation fee of £1.20 (Prolific pays participants in UK currency), while Japanese ones received ¥150. These fees were based on platform guidelines.

---

[4] In Japan we carried out two waves of data collection and stopped the second wave after a full day passed with no new participants. We had already collected data from Lancers for three earlier and related experiments, so we may have exhausted the participant pool.

### Cross-modal discounting task.

We adapted the random-order delayed compensation method (DCM) described by Read et al. (2023). The experiment involved two consumer goods: a Parker pen and a box of Godiva chocolates. These were selected for comparability with earlier studies, which had used similar chocolates and pens because they were familiar goods, had similar retail prices (around $30), and represented one "hedonic" and one "utilitarian" good. They were ideal for the present study because they were familiar in both countries, and the retail prices were also similar in both countries. The pen differed from the one used by Read et al. (2023) because while Parker is well known in both Japan and the USA, Lamy pens (offered in the earlier American studies) are unknown in Japan.

The uni-modal conditions were: a pen today versus a pen in 60 days (*PePe* condition), and chocolates today versus chocolates in 60 days (*ChCh* condition). The cross-modal conditions were: a pen today versus chocolates *in* 60 days (*PeCh* condition), and chocolates today versus a pen in 60 days (*ChPe* condition). In Experiment 1, which used a between-participants design, each participant was assigned to one of these four conditions. In Experiment 2, participants completed all four conditions, presented in a random order.

Participants in each condition made a series of choices between the earlier and later option. In addition to the consumer good, for each choice, one option came with a monetary payment, and the other came with $0/¥0 (in one choice, both options came with zero, to establish a baseline preference between the timed goods). The monetary payment was delayed until after the later outcome could be received, to rule out arbitrage opportunities in which participants could receive enough money to purchase a good before it was provided by the experimenter, or even to receive both goods (see Cubitt et al., 2018). By analyzing these questions we could pinpoint, within a narrow range, the monetary payment just sufficient to make a participant indifferent between the options, and we refer to this amount as the "compensation" for that comparison.

The monetary payments for the U.S. sample were: {$50, $45, $40, $35, $30, $25, $20, $18, $16, $14, $12, $10, $8, $6, $5, $4, $3, $2, $1, $0}. These were the same payment amounts as in Read et al. (2018). The Japanese sample saw corresponding payments in yen: {¥5000, ¥4500, ¥4000, ¥3500, ¥3000, ¥2500, ¥2000, ¥1800, ¥1600, ¥1400, ¥1200, ¥1000, ¥800, ¥600, ¥500, ¥400, ¥300, ¥200, ¥100, ¥0}. We chose the Japanese values to have the same ratios between successive amounts as in the U.S. and to also have the same leading digits (and zeroes thereafter). Fortunately, based on PPP (purchasing power parity), ¥100 is close in value to $1 (OECD, 2025) and this equivalence has also been used in prior literature (e.g., Ishii et al., 2017). In fact, the Japanese Yen is worth a little less than $0.01 (in PPP terms) meaning that a given compensation amount for the Japanese sample is worth less than its converted equivalent for the U.S. sample. This makes our estimates of impatience *conservative* for the Japanese sample, since the Japanese currency is worth less and, by the magnitude effect (one of the most robust findings in intertemporal choice, e.g., Thaler, 1981), people are less patient for trade-offs involving smaller amounts. That is, the magnitude effect predicts less patience in the Japanese sample given our conversion rates. Moreover, if we switch from PPP to exchange rate, this effect is magnified since the exchange rate is even more in favor of the U.S. currency – one dollar exchanges for 160 yen.

**Figure 1**

*Screen shot showing a typical choice in Experiment 1. This is from the chocolate-pen (ChPe) condition in the U.S.*

## Choose between:

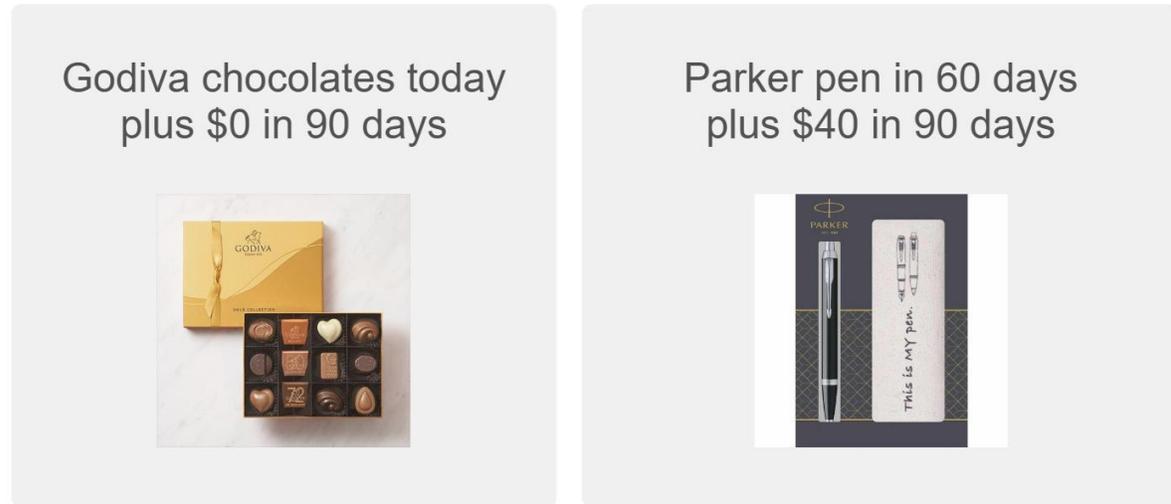

In the two "extreme" questions the monetary payment was $90 (¥9000), either with the earlier or later option.

### Delay discounting/impatience

We included two additional measures of delay discounting. The first comes from Kirby et al. (1999) and consists of 27 monetary choices between a smaller-sooner (SS) and a larger-later (LL) option. A typical choice is between $80 today and $85 in 157 days. The SS option always comes today, the LL option following some days. We call this the *Kirby task*. It is widely used with recent examples being Keidel et al. (2024; 2025) and Eivazi et al. (2022). In Japan, the items were presented in Yen, with the dollar amounts multiplied by 100, using the logic just described. We also used the method of Kirby et al. to estimate the discount rates $k$ from the questions using a one-parameter hyperbolic discounting model: $V = A/(1 + kD)$, where $V$ is the present value of a deferred reward ($A$), determined by delay ($D$), and a free parameter $k$ which is the discount rate.

Additionally, we asked the Preference for Earlier versus Later Income (PELI) question from Burro et al., (2022). In the PELI participants choose between receiving an amount equivalent to their normal monthly income today or twice that amount in one year, with waiting coded as 1 and not waiting as 0. Burro et al. recommended this measure for cross-cultural studies since it compensates for differences in income and currencies without requiring any adjustment.

The correlation between the PELI and $k$ derived from the Kirby task was -0.34 ($p < .001$ by Fisher exact test), suggesting that both measure the same underlying construct. (The negative correlation is intuitive, since those with higher $k$ are less likely to wait in the PELI question.) The correlation between the PELI and the proportion of times the LL option was chosen in the Kirby task was 0.57 ($p < .001$).

**Signed Compensation**

As defined previously, compensation is the amount of money which, if it accompanies the dispreferred of two options, is just sufficient to induce indifference between them. Signed compensation is equal in magnitude to that compensation, but signed as positive (negative) if the sooner (later) option is preferred when neither option has an accompanying payment. Signed compensation therefore measures preference for the sooner option.

To estimate signed compensation, we set out the choice pattern expected for each possible indifference point implied by the monetary payments presented in the study. For example, a person indifferent between good A today (plus $0) and good B in 30 days plus $37.50 would be expected to choose good A today for all monetary payments accompanying good A, or for any amount up to $35 accompanying good B. For payments accompanying good B of $40 or more this person would then choose good B. We set up exemplar choice patterns like this for every indifference point that could be captured with our design. Within each individual's responses, we identified which exemplar choice pattern was most consistent with their decisions in each comparison, thereby determining their most supported indifference points. If more than one pattern was equally well supported, we used the arithmetic mean of the indifference points. See Read et al. (2023) for more details.

**Results**

### *Signed compensation*

Figure 2 presents means and 95% confidence intervals for the signed compensation in both countries. It also presents the difference in signed compensation between treatments, with 95% confidence intervals. A cross-modal effect was observed, as indicated by higher signed compensation for uni-modal than the cross-modal condition. As Figure 2 makes clear, while the degree of discounting is higher in the USA than Japan, the cross-modal effect (the "difference") was the same for both countries.

**Figure 2**

*Signed Compensation for the U.S. sample (left) and the Japanese sample (right) in Experiment 1*

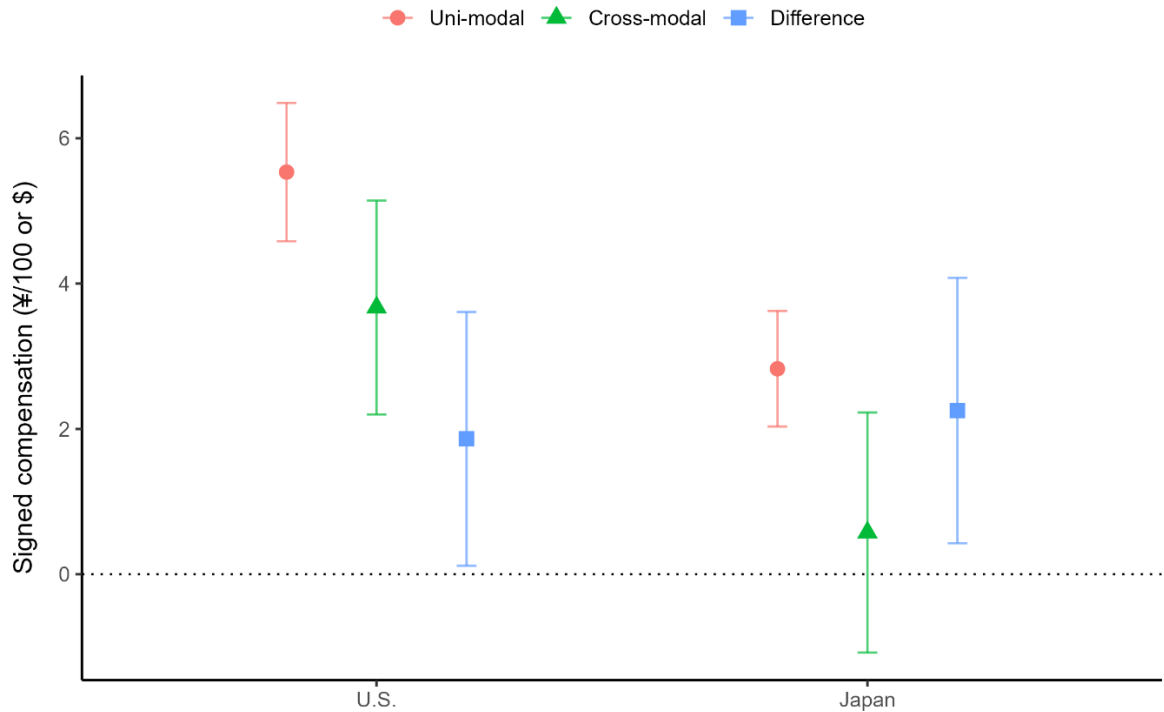

Table 1 presents the results of Ordinary Least Squares (OLS) regressions, with currency values taken from the experiment (USD at face value and Yen divided by 100). Model 1 pools over the entire sample. There was a reliable cross-modal effect with average compensation $2.08 (¥208) lower in the cross-modal than the uni-modal condition. Model 2 adds an indicator for Japan (0,1), a cross-modal indicator, and their interaction. The main effect of cross-modal remains negative and significant, the indicator for Japan is also negative and significant, with Japanese participants demanding significantly less compensation for delay than the U.S. ones. The interaction is not significant, confirming what was evident from Figure 2, that the cross-modal effect is the same across countries.

The remaining models include all other predictors but differ only in the time preference measure included: $k$ (Model 3), the PELI (Model 4), or both (Model 5). We report all three models because, when entered separately, both time preference measures significantly predicted compensation ($p < .001$). However, only the PELI retained significance when both measures were included, indicating that it explains variance beyond that accounted for by $k$.

In addition to analyses based on the currency amounts from the experiment, we conducted parallel analyses using U.S. and Japanese compensation standardized within countries so that both countries had a mean of 0 and a standard deviation of 1. This tests whether the cross-modal effect is the same size in the two countries, holding the absolute level of discounting constant. These analyses are reported in the appendix. The headline result is that there is no appreciable difference between the standardized and non-standardized analyses, further underlining how there are no country differences in the cross-modal effect.

**Table 1**

*OLS regression analysis of Signed Compensation (Experiment 1) using raw data (assuming ￥100 = $1)*

|  | (1) | (2) | (3) | (4) | (5) |
|---|---|---|---|---|---|
| Cross-modal | -2.075** | -1.864* | -1.793* | -1.740* | -1.778* |
|  | (0.646) | (0.888) | (0.887) | (0.886) | (0.886) |
| Japan |  | -2.707** | -2.592** | -2.413* | -2.206* |
|  |  | (0.906) | (0.938) | (0.942) | (0.949) |
| Cross-modal × Japan |  | -0.390 | -0.528 | -0.574 | -0.552 |
|  |  | (1.286) | (1.284) | (1.283) | (1.282) |
| $k$ |  |  | 16.327** |  | 11.611 |
|  |  |  | (6.233) |  | (6.494) |
| PELI |  |  |  | -2.131** | -1.772* |
|  |  |  |  | (0.671) | (0.700) |
| Female |  |  | 0.777 | 0.667 | 0.747 |
|  |  |  | (0.649) | (0.647) | (0.648) |
| Age in Years |  |  | 0.068* | 0.059* | 0.064* |
|  |  |  | (0.027) | (0.027) | (0.027) |
| College Education (0,1) |  |  | -1.368 | -1.258 | -1.234 |
|  |  |  | (0.777) | (0.778) | (0.778) |
| Married (0,1) |  |  | -0.708 | -0.623 | -0.616 |
|  |  |  | (0.671) | (0.671) | (0.671) |
| Employed (0,1) |  |  | 0.234 | 0.300 | 0.289 |
|  |  |  | (0.714) | (0.714) | (0.713) |
| Constant | 4.254*** | 5.535*** | 3.130* | 4.855** | 4.014* |
|  | (0.455) | (0.623) | (1.529) | (1.494) | (1.565) |
| N | 1469 | 1469 | 1468 | 1468 | 1468 |

*Note:* Standard errors are reported in parentheses. *, ** and *** stand for statistical significance at the 5%, 1% and 0.1% level respectivel

### *Discounting between countries*

People from Japan were more patient. Using our conversion rate of $1 = ¥100, the Japanese sample required lower compensation than the U.S. sample for both uni-modal (JP: ¥283 [=$2.83] vs. US: $5.53) and cross-modal (JP: ¥57 [=$0.57] vs. US: $3.67) conditions. Welch's *t* tests indicated these differences were significant: uni-modal, $t(726.8) = -4.29$, $p < .001$, 95% CI [$-3.94$, $-1.47$]; cross-modal, $t(710.6) = -2.75$, $p = .006$, 95% CI [$-5.31$, $-0.8$][8].

The Japanese participants also displayed greater patience for both delay discounting measures. The *k* parameter was significantly lower in Japan ($M = 0.009$) than in the U.S. ($M = 0.038$), $t(1016.1) = -11.16$, $p < .001$, 95% CI [$-0.034$, $-0.024$]; Wilcoxon rank-sum test: $z = -16.22$, $p < .001$. Moreover, a much higher proportion of Japanese participants chose later over earlier income in the PELI (68.6% versus 41.8%), $z = 10.31$, $p < .001$, 95% CI [$0.22$, $0.32$]. Using all measures of time preference, therefore, Japanese participants were significantly more patient than American ones.

### Experiment 2

In Experiment 2, we adapted the random-order DCM for a within-participants design, with everyone taking part in all four conditions. In this way we could measure the cross-modal effect at the individual level, and therefore to explore for the first time whether it differs with individual characteristics. It also provided a further test of whether the cross-modal effect is the same between the two countries and therefore plausibly the result of universal, low-level cognitive processes.

---

[8] As mentioned earlier each Japanese Yen is worth a little less than one American cent. This strengthens the conclusion that the Japanese participants are more patient.

# Method

## *Participants*

Because each participant took part in four treatments, we recruited approximately one-quarter as many participants as in Experiment 1. We invited 210 American participants from Prolific and 210 Japanese participants from Lancers[9]. We maintained a gender balance across samples. Data from two American and one Japanese participant were excluded because they did not complete the study. A further three Japanese participants were excluded for having consistency scores below 0.5. The final sample included 208 American (50% female; $M$ = 35.4 years, *range* = 18–74) and 206 Japanese participants (50% female; $M$ = 42.0 years, *range* = 19–76). On average, the Japanese sample answered the Cognitive Reflection Test (CRT; Frederick, 2005, Harada et al., 2018) questions more successfully (scoring 1.64 versus 1.38); $t(410.59)$ = –2.26, $p$ =0.024, 95% CI [–0.49, –0.03]. This was despite Japanese participants being *less* likely to have a college degree (74% versus 82%); $t(403.11)$ = 2.07, $p$ = .039, 95% CI [.004, .16]). American participants received £3, and Japanese participants received ¥460.

## *Design and Procedure.*

Every participant saw all four conditions (two uni-modal and two cross-modal) in a random order. To ensure understanding, participants were required to correctly answer a comprehension question before each condition. These questions presented participants with an example in which the earlier option came with a $0.50 payment, and a multiple-choice option. For instance, they might be shown a choice between a Pen today (plus a $0.50 payment) and a Pen in 60 days (plus a $0 payment). They would then attempt to choose the correct option from a list of four statements describing what would happen if the earlier option was chosen. The correct answer would be "You would get the Parker pen today and $0.50 in 90 days."

---

[9] Using effect size estimates from the two pretests we calculated that a minimum of 175 participants per group would be needed to detect a significant treatment effect with adequate power. We slightly oversampled to account for potential exclusions.

A notable change from Experiment 1 is the smaller number of choices per condition. We expected 40 choices per condition (160 in total plus four comprehension questions) would be overwhelming, so we reduced it to 22 questions covering the same payment range. These involved 90-day delayed payments of: {$40, $30, $20, $15, $10, $8, $6, $4, $2, $1, $0} for the American sample and {¥4000, ¥3000, ¥2000, ¥1500, ¥1000, ¥800, ¥600, ¥400, ¥200, ¥100, ¥0} for the Japanese sample. All non-zero amounts appeared once with the sooner and once with the later good.

**Figure 3**

*Screen shot showing a typical choice in Experiment 2. This is from the chocolate-pen (ChPe) condition in the U.S.*

| Godiva chocolates | today |
|---|---|
| $0 | in 90 days |
| ○ | |

| Parker pen | in 60 days |
|---|---|
| $6 | in 90 days |
| ○ | |

To explore individual-level factors that may be associated with discounting and especially the cross-modal effect we incorporated additional measures. Cognitive ability was measured with the Cognitive Reflection Test (CRT; Frederick, 2005, Japanese version from Harada et al., 2018). We assessed subjective social status with the McArthur Scale (Adler et al., 2000), in which participants rate their own social status on a scale from 1 (lowest status) to 10 (highest status). Both these measures have been associated with intertemporal decision making (e.g., Ackert et al., 2020; de Wit et al., 2007; Kirby & Maraković, 1995; Reimers et al., 2009). Higher subjective social status has been found to positively correlate with patience in both Japan and the U.S. (Ishii et al., 2017). We also elicited the desirability of chocolates and pens, since this could plausibly influence the weight placed on the outcomes in the choice tasks. Participants separately rated the desirability of each item on a 7-point scale, and we took the difference in these scores to indicate the relative desirability of the chocolates.

**Results**

*Cross-modal effect*

Figure 4 presents means and 95% confidence intervals for the signed compensation, as well as for the difference in signed compensation between treatments, in both countries. As in Experiment 1, a cross-modal effect was observed, as indicated by higher signed compensation for uni-modal than the cross-modal condition. And, again, while the degree of discounting is higher in the USA than Japan, the cross-modal effect did not differ between countries.

**Figure 4**

*Signed Compensation for the U.S. sample (left) and the Japanese sample (right) in Experiment 2*

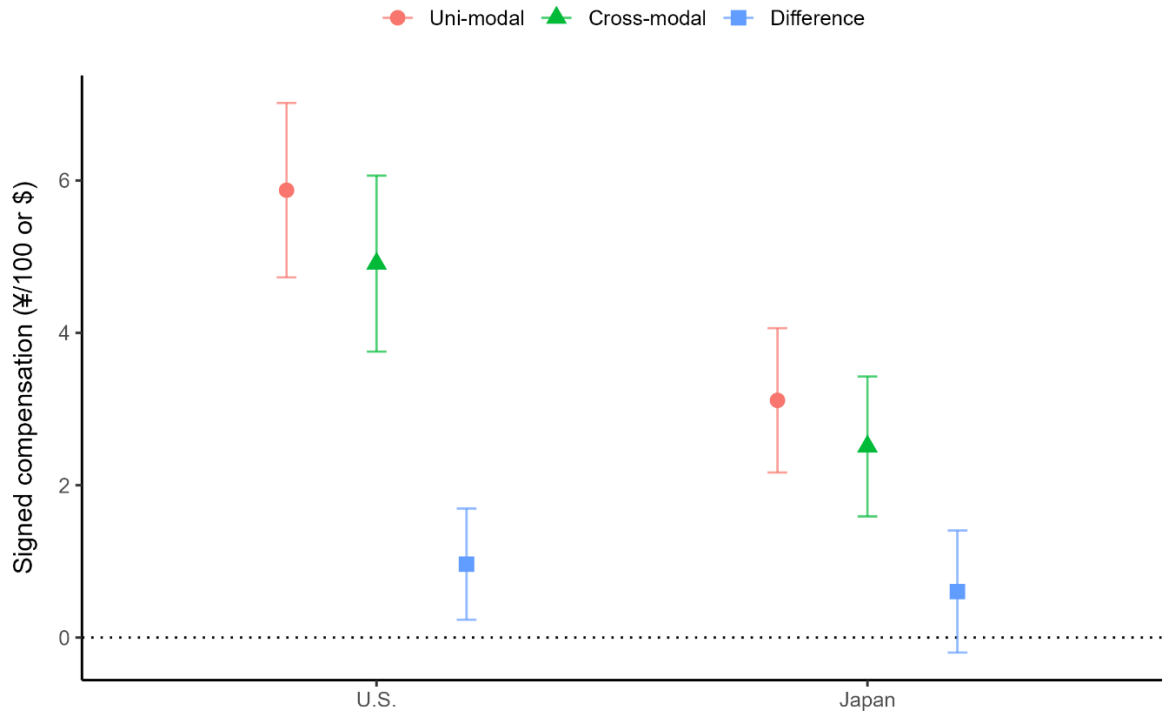

There was a significant overall cross-modal effect. Pooling all data and assuming \$1 = ¥100, the average signed compensation was 4.50 (95% CI [3.74, 5.26]) for the uni-modal condition and 3.71 (95% CI [2.97, 4.46]) for cross-modal, $t(413) = 2.84$, $p = .0047$.

When analyzed separately by country, a similar pattern emerged. In the U.S., the mean signed compensation was higher in the uni-modal (M = \$5.87) than in the cross-modal condition (M = \$4.91), a significant paired difference, $t(207) = 2.59$, $p = .010$, 95% CI [\$0.23, \$1.70]. In the Japanese sample, the mean signed compensation was ¥311 (uni-modal) and ¥251 (cross-modal); although this paired difference was not significant, $t(205) = 1.48$, $p = .141$, 95% CI [–¥20, ¥141].

The direction of the effect is consistent across countries. Furthermore, the two independent pretests conducted with Japanese participants using the same within-participants design both found significant cross-modal effects (reported in Online Appendix A). Overall, the evidence of a within-participants cross-modal effect in Japan is very strong. Lastly, cross-country comparisons revealed that American participants required more compensation than Japanese participants in both conditions: uni-modal (US: $M = \$5.87$; JP: $M = ¥311 [=\$3.11]$), Welch's $t(398.7) = 3.64$, $p < .001$, 95% CI [$1.27, $4.25]; and cross-modal (US: $M = \$4.91$; JP: $M = ¥251 [=\$2.51]$), Welch's $t(392.9) = 3.19$, $p = .0015$, 95% CI [$0.92, $3.88].

Table 3 presents the OLS regressions with standard errors clustered at the participant level to account for the within-participants design. This clustering adjusts for any correlation among observations from the same participant, providing more reliable inferences.

Model 1 includes only the experimental condition. and confirms a cross-modal effect. Signed compensation is on average, $0.79 lower for the cross-modal condition. Model 2 adds country and the country by condition interaction, finding a significant effect of country but no interaction, indicating the cross-modal effect does not differ between the U.S. and Japan. Models 3 to 5 included all other predictors. As in Experiment 1, both $k$ and the PELI are significant predictors of signed compensation, but only the PELI remains when both are included.

### Discounting between countries

Just as in Experiment 1, Japanese participants displayed greater patience than American ones. The $k$ parameter was lower in Japan than the United States ($M = 0.012$ vs $0.025$), $t(409.67) = 3.32$, $p = .001$, 95% CI [0.005, 0.021]; Wilcoxon rank-sum test: $z = 8.17$, $p < .001$. Moreover, significantly more Japanese than American participants selected the patient PELI option (72.3% vs 41.8%), $t(408.9) = -6.58$, $p < .001$, 95% CI [–0.40, –0.21]. Wilcoxon Rank Sum $z = –6.26$, $p < .001$. The correlation between PELI and $k$ was -0.37 ($p < .001$ by Fisher exact test), and that between PELI and the average number of LL choices in the Kirby task was 0.57 ($p < .001$).

Table 2

*OLS regression analysis of Signed Compensation (Experiment 2) using raw data (assuming ￥100 = $1)*

| | (1) | (2) | (3) | (4) | (5) |
|---|---|---|---|---|---|
| Cross-modal | -0.785** | -0.964** | -0.964* | -0.964* | -0.964* |
| | (0.276) | (0.372) | (0.373) | (0.373) | (0.373) |
| Japan | | -2.759*** | -2.245** | -1.592* | -1.496 |
| | | (0.758) | (0.810) | (0.773) | (0.776) |
| Cross-modal × Japan | | 0.359 | 0.359 | 0.359 | 0.359 |
| | | (0.553) | (0.554) | (0.554) | (0.554) |
| CRT | | | -0.036 | -0.139 | 0.018 |
| | | | (0.346) | (0.335) | (0.338) |
| $k$ | | | 47.668** | | 36.865 |
| | | | (18.410) | | (18.943) |
| PELI | | | | -3.834*** | -2.789*** |
| | | | | (0.752) | (0.704) |
| Social status | | | -0.102 | 0.010 | -0.045 |
| | | | (0.231) | (0.229) | (0.226) |
| Female | | | -0.519 | -0.699 | -0.465 |
| | | | (0.764) | (0.771) | (0.746) |
| Age in Years | | | -0.001 | -0.005 | -0.003 |
| | | | (0.026) | (0.025) | (0.025) |
| College education (0,1) | | | -0.832 | -0.854 | -0.717 |
| | | | (1.095) | (1.065) | (1.064) |
| Employed (0,1) | | | 0.079 | -0.106 | -0.003 |
| | | | (0.670) | (0.698) | (0.659) |
| Constant | 4.500*** | 5.872*** | 6.149*** | 8.903*** | 7.222*** |
| | (0.385) | (0.584) | (1.706) | (1.775) | (1.744) |
| N | 1656 | 1656 | 1656 | 1656 | 1656 |

*Note:* Standard errors clustered at the participant level are reported in parentheses. *, ** and *** stand for statistical significance at the 5%, 1% and 0.1% level respectively.

### *Individual level analysis*

We began by categorizing each participant's responses into three patterns: (1) cross-modal effect – compensation was higher in uni-modal than in cross-modal; (2) reverse cross-modal effect – compensation was higher in cross-modal than in uni-modal; and (3) no difference – compensation was the same in the two conditions. These classifications are given in Table 3, which also includes data for the two pretests described in Online Appendix. As demonstrated in earlier papers, according to standard economic assumptions when combined with a concave monetary utility function, the *reverse* cross-modal effect will be most common. Our results so far suggest that the cross-modal effect would be most common, but this study is the first to enable us to test the hypothesis directly.

We found that the cross-modal effect was indeed the most common response pattern, particularly much more common than the reverse. We conducted a one-sample test of proportions using only the observations that exhibited either a cross-modal or a reverse cross-modal pattern (excluding those with no difference). The results confirmed that the proportion of cross-modal patterns was greater in both Japan ($z = 4.47$, $p < .001$, 95% CI [0.60, 0.74]) and the United States ($z = 2.92$, $p = .004$, 95% CI [0.54, 0.68]). This also replicates in the additional Japanese pretests reported in the Appendix and in Table 3.

**Table 3**

*Distribution of response types in Experiment 2 and in the two Pretests*

| Experiment | Cross-Modal | Reverse Cross-Modal | No Difference | Total |
|---|---|---|---|---|
| Experiment 2: Japanese | 113 | 55 | 38 | 206 |
| | 54% | 26% | 18% | |
| Experiment 2: American | 109 | 70 | 29 | 208 |
| | 52% | 34% | 14% | |
| Pretest 1: Japanese | 72 | 29 | 19 | 120 |
| | 60% | 24% | 16% | |
| Pretest 2: Japanese | 76 | 22 | 20 | 118 |
| | 64% | 19% | 17% | |

Next, we explore the individual-level cross-modal effect, which we measured as the difference between the individual uni-modal and cross-modal averages. Consistent with our previous analysis, the country coefficient for Japan is not significant (model 1, Table 4). Model 2 included all other predictors. As in the pooled-data regressions reported above, we also ran specifications that varied depending on which monetary intertemporal choice task was included—$k$, the PELI, or both. However, none of these variables were significantly associated with the cross-modal effect, so we report only the version with both measures included.

While this is by no means a comprehensive investigation, it does suggest that, as proposed by the attentional dilution hypothesis, the cross-modal effect is a universal characteristic which operates prior to factors that might be influenced by education or individual circumstances, such as CRT, subjective social status, or time preference.

**Table 4**

*Individual-level regression analysis of the Cross-modal Effect (Experiment 2)*

|  | (1) | (2) |
|---|---|---|
| Japan | -0.359 | -0.040 |
|  | (0.553) | (0.622) |
| Chocolate desirability |  | -0.183 |
|  |  | (0.130) |
| CRT |  | -0.030 |
|  |  | (0.251) |
| *k* |  | -1.470 |
|  |  | (7.359) |
| PELI |  | -0.626 |
|  |  | (0.630) |
| Social status |  | 0.043 |
|  |  | (0.165) |
| Age |  | -0.025 |
|  |  | (0.024) |
| Female |  | -0.097 |
|  |  | (0.588) |
| College education (0,1) |  | -1.381[*] |
|  |  | (0.697) |
| Employed (0,1) |  | 0.159 |
|  |  | (0.572) |
| Constant | 0.963[*] | 3.196[*] |
|  | (0.390) | (1.421) |
| N | 414 | 414 |

*Note:* Standard errors are reported in parentheses. *, ** and *** stand for statistical significance at the 5%, 1% and 0.1% level respectively.

## General Discussion

Time matters less when outcomes differ, both in America and Japan. We established this through two experiments testing the magnitude of the cross-modal effect in the WEIRD (Western, Educated, Industrialized, Rich, and Democratic, Henrich et al. 2010) context of the United States and in the less Westernized, and thus less WEIRD, context of Japan. We make three key contributions: first, we replicated the cross-modal effect in the American setting in which it was originally documented and also replicated previous assertions that Japanese people are more patient than Americans; second, we provided the first cross-cultural comparison of the cross-modal effect, finding no difference between Japan and the USA; third, we found evidence suggesting the cross-modal effect is a cognitive universal, neither influenced by culture nor by individual differences such as cognitive ability, subjective social status or even age and gender, despite differences in patience between the cultural settings we studied.

The hypothesized attentional dilution mechanism, rooted as it is in low-level cognitive processes, is given substantial support by the universality of the cross-modal effect. It asserts that time (as well as other attributes) is given more decision weight the more attention is directed to it, and that the amount of attention directed toward time can be decreased by increasing the complexity of the task. This adds to much recent work showing the link between the allocation of attention and decision weights (Franco-Watkins et al., 2016; Keidel et al., 2024; Pachur et al., 2018; Yechiam & Hochman, 2013). As well as increasing our understanding of how attention influences intertemporal choice, our results further support the conclusion that the *behavioral* discounting of a given outcome does not arise from a fixed individual parameter such as a pure rate of time preference but is a highly context-sensitive process (Lempert & Phelps, 2016; Reeck et al., 2017; Scholten et al., 2024).

Our findings are also consistent with previous work (e.g., Graham & Sano, 1984; Hofstede insights, n.d.; Levine, 2008; Ishii et al., 2017) showing that Japanese participants are more patient than their American counterparts. We observed this in decisions about receiving goods at different times, as well as in two separate monetary intertemporal choice tasks. This increases the evidence that despite universalities like the cross-modal effect, there remain systematic cultural differences in economic preferences (e.g., Burro et al., 2022; Falk et al., 2018; Ruggeri et al., 2022; Vieider, 2019; Wang et al., 2016).

Given the difference in overall patience, it was plausible that the cross-modal effect would also differ between the two cultural settings. For instance, the effect might be smaller in Japan because less weight is placed on delay in intertemporal choices in Japan than in the U.S. However, the effect was highly consistent across the two countries. These contrasting results – universality of the cross-modal effect and large cultural variations in time discounting – are relevant to broader debates about the generalizability of psychological findings beyond WEIRD samples (Henrich  et al., 2010). Many decision-making anomalies vary greatly from country to country. Delay discounting itself is one of these, with the excessive discounting that constitutes a major puzzle in North American samples being much less prevalent in Japan. But when findings are seen in different cultures, as shown by Yoon et al. (2019) in the context of anchoring, it suggests they are underpinned by something universal.

As well as replicating the cross-modal effect in the U.S. and documenting it in another setting, we also examined whether the within-participants cross-modal effect differs according to individual characteristics. We found that approximately twice as many people showed a cross-modal effect as the reverse inconsistency. Interestingly, we found no evidence that the strength of the cross-modal effect depends on any of the individual characteristics we measured, characteristics that are known to be closely related to time discounting itself.

We also further examined the underlying processes involved in intertemporal decision making. The degree of discounting in cross-modal decisions depends on whether the experimental design is within- or between-participants. In Experiment 1, we employed a between-participants design, so those in the uni-modal conditions never faced cross-modal choices, and vice versa. The within-participants design of Experiment 2, however, ensured participants saw all conditions. Attentional dilution predicts that if time is made more salient, as it would be by mixing both cross-modal and uni-modal decisions, then time will have a greater impact on cross-modal decisions. The top four rows of Table 5 show the mean cross-modal and uni-modal compensation for Experiments 1 and 2. They indicate that whereas the uni-modal compensation is not altered by the within-participants design, the cross-modal compensation is, with less discounting in the within-participants setting.

We formally tested this with a meta-analysis that combined Experiments 1 and 2, reported in full in Appendix B. We conducted a one-stage individual participant data mixed-effects model pooling both experiments (c.f., Tamási et al., 2022). The model included an interaction between Modality and Design; a random intercept for participant; cluster-robust standard errors, with N = 3,125 observations from 1,883 participants. Using the raw dependent variable (signed compensation), the interaction showed a substantial attenuation of the cross-modal effect (about 62%), with estimates of -0.785 in the within-participants study versus -2.075 in the between-subjects study (b = +1.289, SE = 0.705, p = .067) although the difference was marginally significant. A model correcting for skewness in the data suggested the effect was larger and statistically significant[10]. While this conclusion is not definitive, it appears likely that the cross-modal effect is substantially smaller in the within-participants design, which fits with previous literature showing that attributes given little weight between-participants become more important within-participants (Charness et al., 2012; Hsee et al., 1999; Keren & Raaijmakers, 1988).

Table 5 summarizes all studies we have conducted using the same or related designs. Along with those described in this paper are those published earlier (Cubitt et al., 2018; Read et al., 2023), and two that were conducted for an earlier project (Cubitt et al., 2018) but as yet unpublished[11].

---

[10] We analyzed the data using an inverse-hyperbolic-sine transformation, which functions similarly to a log transform but is suitable for highly skewed non-normal data where values are 0 and negative. This result was statistically reliable (b = +0.223, SE = 0.086, p = .009), indicating approximately 48% attenuation of the transformed effect in the within-participant condition (within: -0.244; between: -0.467).

[11] The journal editors judged these studies would be better left out because they were mere replications – we suspect a journal editor would likely not make that request nowadays.

There are a few things to note here. Firstly, the cross-modal effect is very reliable. We have not carried out an experiment in which we did not observe it. Secondly, the between- and within-participants difference is also striking, since earlier studies based on between-participants designs show much greater effects than within-participants designs.

It is worth mentioning, as well, that earlier studies, all conducted on U.S. samples, show a larger cross-modal effect than do those reported in this paper. While all the studies in the table use closely related designs, there are several differences. One noteworthy one is that the goods offered in the new studies are not quite as luxurious as those in the earlier ones, and another is that there are some differences in the methods used. In particular, the studies conducted for the Cubitt et al. (2018) paper first asked people to choose between options offering no payment, before filtering them into a one-sided choice list based on their initial decision, offering money amounts with their less preferred option to identify indifference, while later studies (like the ones presented here) elicited decisions for all the items on a double-sided choice list with money amounts appearing on both sides. While this is not the place to examine these differences in detail, further research might usefully examine moderators of the cross-modal effect, and differences in how the effect is measured and in the goods themselves may form one line of enquiry. The principle of the attention-dilution mechanism is that any factors that shift attention between different features of the choice task will have a corresponding and predictable effect on the decision weight assigned to those features. As such, any experimental manipulation of attention would provide further tests of the robustness and generality of the attentional dilution mechanism in intertemporal choice and beyond.

**Table 5**
*Summary of the Cross Modal Effect across Studies*

| Study | Uni-Modal | Cross-Modal | Difference | Within participants |
|---|---|---|---|---|
| Experiment 1: U.S. | 5.53<br>390 | 3.67<br>378 | 1.86 | No |
| Experiment 1: JP/100 | 2.83<br>350 | 0.57<br>351 | 2.25 | No |
| Experiment 2: U.S. | 5.87<br>208 | 4.91<br>208 | 0.96 | Yes |
| Experiment 2: JP/100 | 3.11<br>206 | 2.51<br>206 | 0.60 | Yes |
| Pretest 1: JP/100 | 3.42<br>120 | 2.36<br>120 | 1.06 | Yes |
| Pretest 2: JP/100 | 2.96<br>118 | 1.67<br>118 | 1.29 | Yes |
| Read et al. (2023): U.S. | 6.07<br>154 | 0.57<br>151 | 5.50 | No |
| Cubitt et al. (2018): U.S. | 4.63<br>151 | 0.46<br>149 | 4.17 | No |
| Cubitt et al. Unpublished: U.S., Pen and chocolate | 8.88<br>210 | 0.265<br>392 | 8.62 | No |
| Cubitt et al. Unpublished: U.S., Camera and headphone | 14.29<br>210 | -5.88<br>392 | 20.2 | No |

**Statements and Declarations**

**Funding**

This research has benefited from the financial support of Grants-in-aid for Scientific Research No. 21K13257 from the Japan Society for the Promotion of Science.

**Conflicts of interest/Competing interests**

The authors declare that they have no conflict of interest.

**Code availability**

The data and code (programmed in R and Stata) to replicate the results of the experiment are available on OSF (https://osf.io/3jdya/overview?view_only=f0ca4fd9e67c448bb08df362cd7166ba).

**Internal Review Board Approval**

The experiments reported in this paper were approved by the ethical committee at Hitotsubashi University (No. 2022D009).